\documentclass[pre,twocolumn,superscriptaddress,showpacs,floatfix]{revtex4}
\usepackage{listings}
\usepackage{fancyvrb}
\usepackage{graphicx}
\usepackage{latexsym}
\usepackage{bm}
\usepackage{natbib}
\usepackage{dcolumn}
\usepackage{amsmath}
\usepackage{amstext}
\usepackage[english]{babel}
\usepackage{flafter}
\usepackage{url}
\usepackage{multirow}
\usepackage{float}
\usepackage[usenames]{color}
\newfloat{figtable}{t}{lop}
\floatname{figtable}{Table}

\begin{document}

\title{Physics peeks into the ballot box}
\author{Santo Fortunato}
\affiliation{Department of Biomedical Engineering and Computational
  Science, School of Science, Aalto University, P.O. Box 12200, FI-00076, Espoo, Finland}
\affiliation{Complex Networks and Systems Lagrange Lab, ISI Foundation, Via
  Alassio 11/C, 10126 Torino, Italy} 
\author{Claudio Castellano}
\affiliation{Istituto dei Sistemi Complessi (ISC-CNR), Via dei Taurini 19,
I-00185 Roma, Italy}
\affiliation{Dipartimento di Fisica, “Sapienza” Universit\`a di Roma,
P.le A. Moro 2, I-00185 Roma, Italy}

\begin{abstract} 
Electoral results show universal features, such as
statistics of candidates' performance and turnout rates,
in different countries and over time.
Are voters as predictable as atoms? 
\end{abstract}

\maketitle

{\bf The idea that humans}, in particular circumstances, might behave
like {\it social atoms}, i.e. they might obey rules that
yield predictable collective patterns, just like atoms and molecules,
dates back many years and has several illustrious fathers,
like political philosopher John Stuart Mill and
social scientists Auguste Comte and Adolphe Quetelet. 
The principle is that, when a great multitude of individuals interact,
the choice of each of them, which in principle is free, is
constrained by the presence of the others, and regular collective behaviors 
may emerge.
In the last two decades this idea has developed into a flourishing
field of investigation, due to the unprecedented availability of large
datasets on social phenomena, and powerful computers to process them. 
Statistical physics, developed to deal with systems composed of many
particles, has naturally provided a conceptual framework for
understanding such systems, even if here nothing less than humans
are the elementary constituents. 
Clearly, such an endeavor does not come without obstacles.
The complexity of individual ``social atoms'' and the intricacy of
their interactions rule out the possibility to
formulate principles and laws as rigorous as in the physics of matter.
Nevertheless, regular patterns in collective social phenomena abound, and
they call for explanations, ideally via simple models.


Works of physicists on elections started to appear in the late 90s.
As of now, many
different types of elections in several contries have been studied and
modelled, and various issues have been addressed. 
In this study we discuss recent results on the competition between
candidates in proportional elections and on the statistics of turnout rates.

\vskip0.4cm

\section{The performance of candidates}

In electoral competitions many candidates compete for one or few seats.
Their success can range from landslide victory to complete failure with
only a handful of votes received. How does this huge variability
come about?

In 1999 Costa Filho et al. studied vote statistics in
Brazilian Parliamentary proportional elections.
In this kind of elections, at odds with what occurs in the US,
in each electoral district, a few to a few dozens seats are awarded
simultaneously, and the number of seats won by a party 
is proportionate to the number of votes received.
Costa Filho et al. considered all candidates, winners and losers, 
regardless of their electoral district or political affiliation
and computed the histogram of the number of votes received by each of them.
In the central range of the number of votes, 
the histogram has a regular profile, where the number of
candidates receiving $v$ votes appears inversely proportional
to $v$. This means that the number of candidates who
obtained 1000 votes, say, is approximately 10 times higher than the
number of candidates who collected 10000 votes.
This type of analysis neglects a factor that is fundamental in the
political life of most countries: the influence of parties. 
A great deal of the electoral success of a candidate depends on the
popularity of the party's political agenda
he/she is running for.
The observed histogram reflects the complex
intertwined effect of the competition among parties and the competition
among different candidates within parties.
Given the peculiarities of the specific political landscape of each
country, one can hardly expect to see similar results across 
different countries.

For this reason, we have proposed a different analysis for
proportional elections where voters select a party and
in addition express a preference for one of the candidates within
the party list. 
To disentangle the role of the party and the role of the candidate,
the key variable is not the absolute candidate performance,
expressed by the number of votes obtained, but the
performance relative to the average performance of competitors
within the same party.
For each party list, we divide the number $v$ of votes collected by each
candidate by the average score $v_0$ of all candidates in the list.
The histogram of
the relative performance $v/v_0$ turns out to be not only very stable
for different elections in the same country, but also across countries
[Fig. 1 (left)].
The generality of this result, and the stability of the performance curve
over more than half a century, suggests that we are facing a basic
feature of social dynamics, independent of the cultural, historical,
economic and technological context where the elections took place.

We model the electoral campaign of candidates as a word-of-mouth
process, starting from the candidates and branching out to the
community of voters, through their social relationships. 
Each candidate convinces his/her immediate contacts with a certain
probability $r$. Only voters who have not yet made up their mind may be convinced to
choose a candidate. Convinced voters, in turn,
become activists and campaign in favor of the candidate they have
chosen. 
The number of contacts of each individual is a random variable $k$,
broadly distributed as $p(k) \sim k^{-\alpha}$ to account for the large
variability in personal inclinations and social stance.

The process stops when each voter has picked a candidate. 
The model yields a distribution of candidates' performance which
matches well the empirical one [Fig. 1 (left)]
and is rather robust with respect to
variations of the parameters $r$ and $\alpha$.
It turns out that the most successful candidates are those who manage to
acquire the largest number of voters in the initial stages of the
process. Indeed, with an initially high number of activists on their
side, they may target and eventually gain a high number of supporters,
thus limiting the opportunities of their competitors.

\section{Turnout rates}

Imitation and adaptation are the main features of the social atom, and
frequently overrule the individual freedom to make decisions. Knowing
how people's decisions are affected by those of their contacts is
fundamental in marketing, policy making, panic situations,
etc. Borghesi et al. have used space-resolved election data to
address this issue.
They focused on the main decision of a voter: shall I vote or not?
Turnout statistics are available at the municipal
level for many countries and years. For each municipality one can
easily extract the geographical location as well. In this way it is
possible to check whether there are spatial correlations between
turnout rates of different towns, i.e. whether the decision of many
people to go to vote (or not to go) in one place is related
to the behavior of the voters in nearby places.

Borghesi et al. studied the logarithm of the ratio between the
number of registered voters who participated in each election and
the number of those who did not. The histogram of this variable,
called logarithmic turnout rate (LTR), is a
skewed bell-shaped curve that, after proper rescaling,
is essentially the same in different countries and elections, for
municipalities having approximately the same population.
The main result of the analysis, however, is that participation to
elections is not independent in different places.
Turnout rates are correlated over long ranges in space [Fig. 1 (right)]:
quantitatively, the correlation function of LTR decays logarithmically
with the distance r between the locations ($C(r) \sim  \ln L/r$), where
$L$ is of the order of the size of the country.

The phenomenology uncovered can be reproduced in terms of a decision model
where a voter actually participates to the election only if his
own propensity to vote overcomes a threshold. The propensity of
each voter is modeled as a continuous unbounded variable, resulting from
the sum of three contributions: an idiosyncratic disposition
(specific to each individual and varying in time), a city-dependent term
(with short-scale spatial correlations) and a slowly varying 
``cultural field''. The latter transports (in space) and keeps the
memory (in time) of the collective intentions and is responsible
for the long-range spatial correlations of turnout rates.
The evolution of this cultural field is affected by two types of mechanisms.
People travel to neighbouring cities and interact with acquaintances,
carrying their own local cultural specificity and exchanging ideas and
beliefs. In so doing, cultural differences between near areas tend
to narrow and this implies that the cultural field undergoes a
diffusion process over geographically separated locations.
The mood of the community in a given area may also be influenced by
special events, like the closing down of a factory, or the construction 
of new infrastructure, which can be modeled by a random noise term.
The cultural field thus obeys a diffusion equation in the presence
of random noise, whose solution, after a transient leading to stationarity,
displays the same spatial correlation profile found in the
electoral data. The fit of the model predictions with the empirical
distributions reveals that the equilibration time of the cultural
field dynamics is very long, of the order of a century, which is
consistent with the observed historical persistence of the voting
habits in different areas of a country.

The findings we have discussed confirm that elections are phenomena
that can be understood quantitatively with the concepts and tools of
physics. This new line of research has just started to scratch the
surface of the wealth of information contained in electoral data.
It promises to shed light on the decisional processes of voters,
with potential strong impact on the policy of governments and parties,
the design of electoral systems and the organization of political
campaigns. 

\section{Additional resources}

\begin{itemize}
\item{M. Buchanan, {\it The social atom}, Bloomsbury, New York (2007)}


\item{C. Castellano, S. Fortunato, V. Loreto, ``Statistical physics of
  social dynamics'', {\it Rev. Mod. Phys.} {\bf 81}, 591 (2009).}
\item{R. N. Costa Filho et al., ``Scaling behavior in a proportional
    voting process'', {\it Phys. Rev. E} {\bf 60}, 1067 (1999).}
\item{S. Fortunato, C. Castellano, ``Scaling and universality in
    proportional elections'', {\it Phys. Rev. Lett.} {\bf 99}, 138701 (2007).}
\item{C. Borghesi, J.-C. Raynal, J.-P. Bouchaud, ``Election Turnout Statistics in Many Countries: Similarities, Differences, and a Diffusive Field Model for Decision-Making'', {\it PLoS ONE} {\bf 7}, e36289 (2012).}

\end{itemize}

\begin{figure*}
\includegraphics[width=8.5cm]{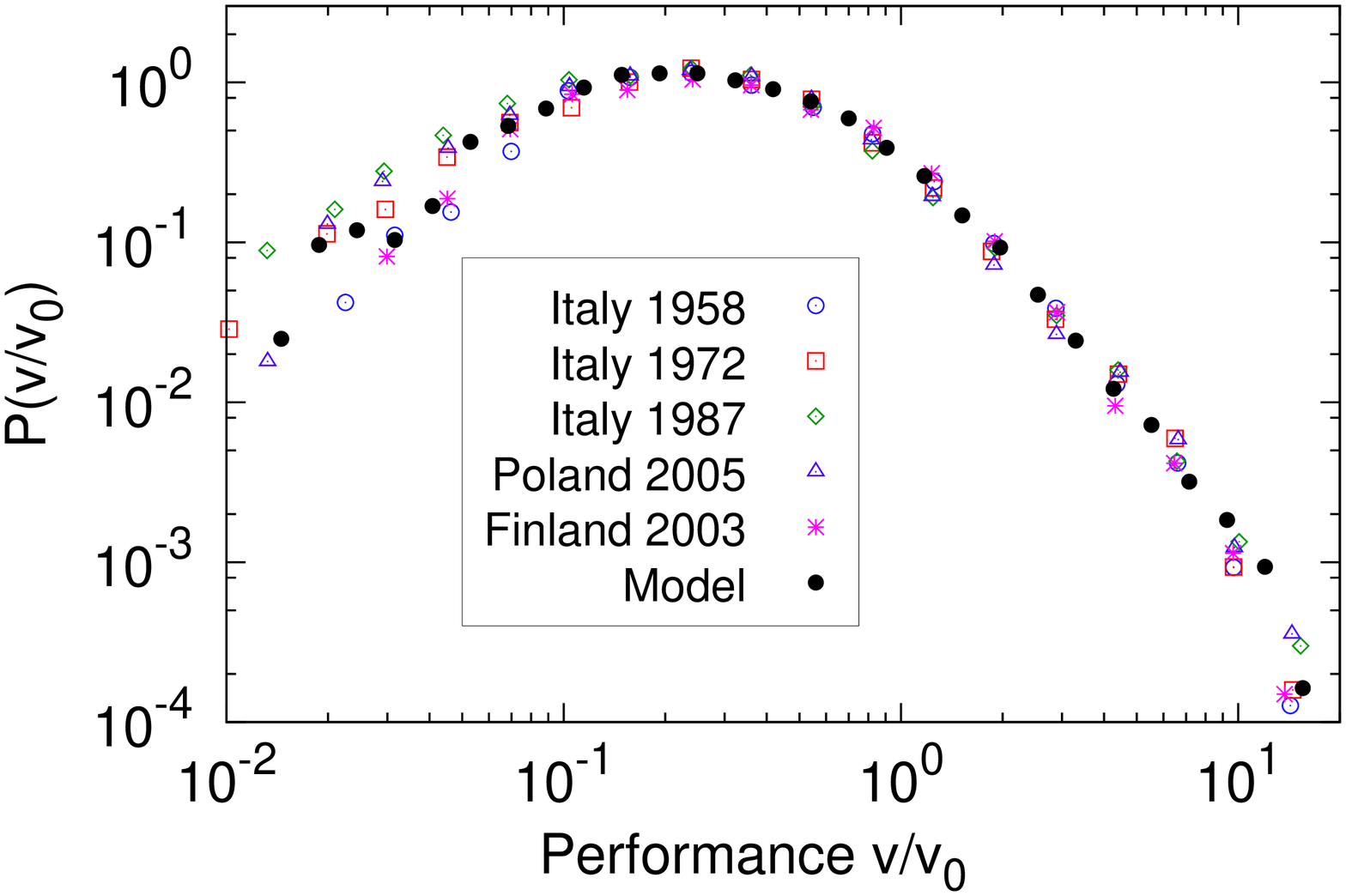}
\includegraphics[width=8.5cm]{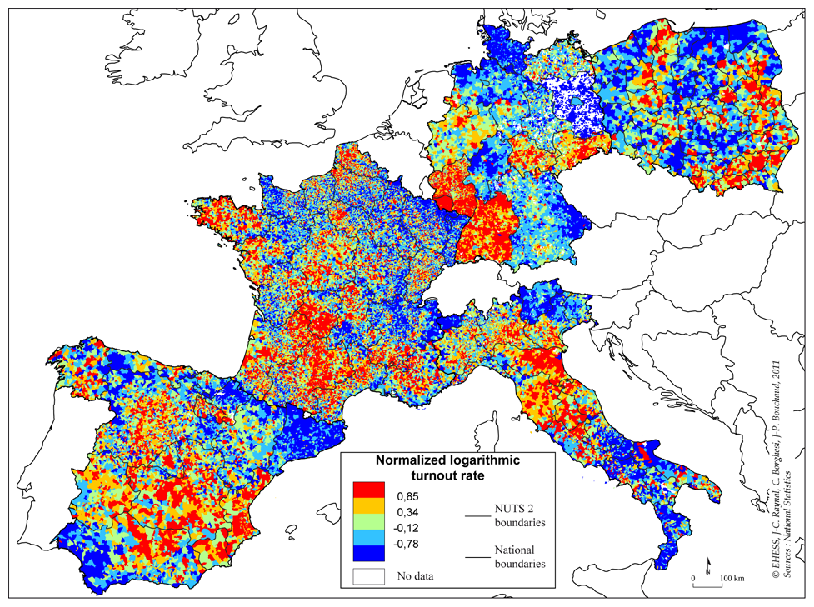}
\label{fig1}
\caption{(Left) Histogram of the performance of a candidate
  with respect to the average score of his/her party competitors in proportional
Parliamentary elections with open lists. The resulting pattern is the same in
different countries and years, hinting to the presence of a simple
underlying mechanism behind the candidates' competition. A model based
on word-of-mouth spreading describes well the universal curve (black dots). 
@ 2007 from the American Physical Society. (Right) Heat
map of the turnout rates for the 2004 European Parliament election in
France, Germany, Italy, Poland and Spain. Red and blue indicate municipalities
with high and low turnout rates, respectively. The pattern is heterogeneous
and displays long-range correlations between different geographical
areas: locations with high (low) turnout are likely to be in
geographical proximity of locations with high (low) turnout.}
\end{figure*}

\end{document}